\documentclass[prb,twocolumn,showpacs,amssymb]{revtex4-1}

\usepackage{graphicx}
\usepackage{dcolumn}
\usepackage{bm}
\usepackage{amssymb}

\begin{document}


\title{Regularized perturbative series for the ionization potential of atomic ions}

\author{Gabriel Gil$^{1,2}$ and Augusto Gonzalez$^1$}
\affiliation{$^1$Institute of Cybernetics, Mathematics and Physics, Havana, Cuba}
\affiliation{$^2$CNR Institute of Nanosciences S3, Modena, Italy}

\begin{abstract}

We study $N$-electron atoms with nuclear charge $Z$. It is well known that, in the cationic ($Z > N$) high-$Z$ region, the atom behaves as a weakly interacting system. The anionic ($Z < N$) regime, on the other hand, is characterized by an instability threshold at $Z_c \lesssim N-1$, below which the atom spontaneously emits an electron. We construct a regularized perturbative series (RPS) for the ionization potential of ions in an isoelectronic sequence that exactly reproduces both, the large $Z$ and the $Z$ near $Z_c$ limits. The large-$Z$ expansion coefficients are analytically computed from perturbation theory, whereas the slope of the energy curve at $Z=N-1$ is computed from a kind of zero-range forces theory that uses as input the electron affinity and the covalent radius of the neutral atom with $N-1$ electrons. Relativistic effects, at the level of first-order perturbation theory, are considered. Our RPS formula is to be used in order to check the consistency of the ionization potential values for atomic ions contained in the NIST database.

\end{abstract}



\maketitle

\section{Introduction}
\label{Introduction}

Since the foundation of Quantum Mechanics, a huge amount of data on energy levels, linewidths and other properties of atoms have been recorded. Very often, such compilations are still waiting for a qualitative analysis, based on simple models. 

In recent articles \cite{pra1,pra2}, on the basis of the scaling suggested by Thomas-Fermi theory \cite{TF}, we have demonstrated universality in the ionization potentials and the correlation energies of atomic ions.

In the present paper, we construct an analytical expression for the ionization energy of atomic ions, which is to be used in order to detect problematic values \cite{NISTrev1,NISTrev2} in the numbers provided by the NIST database.\cite{NIST}

Our expression is a regularized perturbative series (RPS), previously employed in other contexts \cite{renormS}. We use perturbation theory in $1/Z~$ \cite{Z-expansion} in order to compute the first two coefficients of the energy series in the large-$Z$ region. Additionaly, we require our RPS to reproduce the value of the ionization potential at $Z=N-1$ (i.e. the electron affinity) and the slope of the curve at this point. The latter is computed from a kind of zero-range forces theory that uses as input the electron affinity and the covalent radius of the neutral atom with $N-1$ electrons. \cite{ourModPhysLett} The RPS continuosly interpolates between the $Z \approx N-1$ and large-$Z$ limits for a given isoelectronic sequence. 

Hereunder, we provide the main formulae entering the RPS expression. A few isoelectronic sequences are studied in quality of examples. A systematic analysis of the NIST database is contained in Refs. \onlinecite{NISTrev1,NISTrev2}. 

\section{Atoms near the anionic instability threshold}
\label{anionic}

It is well known that, for large $Z$, the attraction of the electrons by the nucleus is stronger than electron-electron repulsion. On the contrary, for the neutral atom both contributions are more or less balanced, and in the anionic domain this balance may even be broken at a given $Z_c \lesssim N-1$, where the atom spontaneously autoionizes.

First-principle calculations \cite{Hogreve} and some extrapolations \cite{Kais} indicate that $Z_c$ is indeed very close to $N-1$, excluding the possibility of doubly charged negative ions. A recent result by Gridnev,\cite{Gridnev} on the other hand, rigorously states that the wave function is normalized at threshold. If we combine this result with perturbation theory, we get that the binding energy exhibits a linear dependence on $Z$ near $Z_c$.

In a previous paper,\cite{ourModPhysLett} we compute the slope of the curve not at $Z_c$, but at $Z=N-1$. At this value of $Z$ the outermost electron weakly interacts with the neutral core and the interaction is short-ranged. It can be shown that conditions are fulfilled for the application of zero-range forces theory. \cite{zero-range} The slope of the curve may be computed from:\cite{ourModPhysLett}  

\begin{equation}
s = 2 \kappa e^{2 \kappa R} \int_{R}^{\infty}{dr e^{-2 \kappa r}/r} ~,
\label{eq1}
\end{equation}

\noindent where $\kappa = \sqrt{ 2 E_a}$ and $E_a$ is the electron affinity of the neutral system with $N-1$ electrons. Atomic units are to be used everywhere in the paper. $R$ is related to the size of the core, containing nuclear charge $Z$ and $N-1$ electrons. For computational purposes, we use the covalent radius of the $N-1$ electron atom as an estimation of $R$.

Eq. (\ref{eq1}) will be used in Sect. \ref{regular}, where we construct a RPS for the ionization potential.

\section{The large-$Z$ limit}
\label{Z_limit}

In the following, we shall construct the large-$Z$ series for the atomic energy. This is, in fact, a formal limit. In nature, atomic ions become unstable for large $Z$, showing a threshold for electron-positron pair production at $Z \sim 137$.\cite{pair-production} Performing the scaling $r_i \rightarrow r_i / Z$ in the non-relativistic Hamiltonian, we get:

\begin{equation}
\hat{H} = Z^2 \left\{ \sum_{i=1}^{N}\frac{\hat{p}^2_i}{2}-\sum_{i=1}^{N}\frac{1}{r_i}+\frac{1}{Z}\sum_{i<j}\frac{1}{|\vec{r}_i-\vec{r}_j|} \right\} .
\label{eq2}
\end{equation}

\noindent Notice that the expression inside brackets has a one-particle contribution (kinetic energy plus nuclear attraction) and the two-particle repulsion between electrons. The latter is of order $1/Z$. At large values of $Z$, the atom can be described as a system of non-interacting electrons in the central Coulomb field of the nucleus. The energy in this leading approximation is:

\begin{equation}
E_0=-\sum_{i=1}^{N}\frac{Z^2}{2 n_i^2} ~.
\label{eq3}
\end{equation}

Next, we shall include electron repulsions in first order perturbation theory. The energy is written as:

\begin{equation}
E= E_0 + Z \left\langle\Psi_0 \left| \sum_{i<j}\frac{1}{|\vec{r}_i - \vec{r}_j|} \right| \Psi_0 \right\rangle ,
\label{eq4}
\end{equation} 

\noindent where $\Psi_0$ is the Slater determinant made of hydrogenic functions. Corrections are explicitly given by:

\begin{eqnarray}
E_1 &=& Z \left\langle\Psi_0 \left| \sum_{i<j}\frac{1}{|\vec{r}_i - \vec{r}_j|} \right| \Psi_0 \right\rangle \nonumber \\
&=& Z \sum_{i<j}^{N}\left\{ \left\langle i j \left| \right| i j \right\rangle - \left\langle i j \left| \right|j i \right\rangle 
\right\} .
\label{eq5}
\end{eqnarray}  

\noindent Note that the sums runs over the occupied orbitals $\left|i\right\rangle$ and $\left|j\right\rangle$ in the Slater determinant, and that $\left\langle i j \left| \right| i j \right\rangle$ and $\left\langle i j \left| \right| j i \right\rangle$ denote, respectively, direct and exchange two-electron Coulomb integrals involving orbitals $i$ and $j$. Their explicit expression can be found in Ref. [\onlinecite{tesis}].

Once we constructed a series for the total energy:

\begin{equation}
E=a_2 Z^2 + a_1 Z + ... ~,
\label{eq6}
\end{equation}

\noindent one can find also a similar expression for the ionization potential, defined as $I_p(N,Z)=E(N-1,Z)-E(N,Z)$. We get:

\begin{equation}
I_p=b_2 Z^2 + b_1 Z + ... ~,
\label{Z_series}
\end{equation}

\noindent where,

\begin{eqnarray}
b_2 &=& \frac{1}{2 n_f^2} \label{b2} ~,\\ 
b_1 &=& -\sum_{j=1}^{N}\left\{\left\langle N j \left| \right| N j \right\rangle - \left\langle N j \left| \right| j N \right\rangle \right\} ~. \label{b1}
\end{eqnarray}

\noindent In these equations, $n_f$ is the principal quantum number of the last electronic shell, 
and $\left|N\right\rangle$ -- the last occupied orbital. 

To end up this section, we shall stress that, in the large-$N$ limit:

\begin{eqnarray}
b_2 &=& \frac{1}{2} \left (\frac{2}{3} \right)^{2/3} N^{-2/3} + ... ~, \label{largeNb2}\\ 
b_1 &=& -0.72~N^{1/3} + ... \label{largeNb1}
\end{eqnarray}

\noindent Eq. (\ref{largeNb2}) comes from analytical estimations, whereas Eq. (\ref{largeNb1}) comes from a fit to the numerical results. These functional forms are consistent with the dependence $I_p \approx Z^2 N^{-2/3} f(N/Z)$, suggested by Thomas-Fermi theory.\cite{pra1}

\subsection{Relativistic corrections}
\label{rel_corr}

At large $Z$, a relativistic approach is required. In the leading approximation, one should solve the Dirac equation for an electron in a central Coulomb field. We choose a simpler approach in which relativistic corrections are computed in first order degenerate perturbation theory,

\begin{eqnarray}
\bar{E}_{\lambda} &=& E^{(0)}_{\lambda} + \left\langle \lambda|V_{rel}|\lambda \right\rangle ~, \label{e_rel}\\
\left|\bar{\lambda}\right\rangle &=& \left|\lambda\right\rangle + \sum_{j \in S_\perp(\lambda)}{\frac{ \left|j\right\rangle \left\langle j|V_{rel}|\lambda \right\rangle }{ E^{(0)}_{\lambda} - E^{(0)}_j } } ~.
\label{wf_rel}
\end{eqnarray}

\noindent Both $\left|\lambda\right \rangle$ and $\left|j \right \rangle$ are eigenstates of the non-relativistic one-electron Hamiltonian. Greek indices label states for which the total angular momentum (orbital plus spin) is a good quantum number. The relativistic perturbation, $V_{rel}$, includes the kinetic ($\sim \hat p^4$), spin-orbit ($\sim \hat{\vec{L}} \cdot \hat{\vec{S}}$), and the Darwin term ($\sim \delta(\vec{r})$) terms.\cite{Bransden} The sum in Eq. (\ref{wf_rel}) runs over the space orthogonal to $\left|\lambda \right \rangle$.

The first order correction to $b_1$ can be obtained by replacing the non-relativistic states $\left|j\right\rangle$ and $\left|N\right\rangle$ in Eq. (\ref{b1}) by the expression (\ref{wf_rel}), yielding:

\begin{widetext}
\begin{eqnarray}
b_1 &=& \sum_{\lambda \neq \sigma} {\left\{ \left\langle \sigma \lambda \left| \right| \sigma \lambda \right\rangle - \left\langle \sigma \lambda \left| \right| \lambda \sigma\right\rangle \right\}}  
+ 2 \sum_{\lambda \neq \sigma}{\sum_{k \in S_{\perp}^{(\sigma)}}{\frac{\left\langle \sigma \lambda \left| \right| k \lambda \right\rangle - \left\langle \sigma \lambda \left| \right| \lambda k \right\rangle}{E^{(0)}_{\sigma}-E^{(0)}_k} \left\langle k|V_{rel}| \sigma\right\rangle}}
\nonumber \\ 
&+& 2 \sum_{\lambda \neq \sigma}{\sum_{k \in S_{\perp}^{(\lambda)}}\frac{\left\langle \sigma \lambda \left| \right| \sigma k \right\rangle - \left\langle \sigma \lambda \left| \right| k \sigma \right\rangle}{E^{(0)}_{\lambda}-E^{(0)}_k} \left\langle k|V_{rel}| \lambda\right\rangle} ~,
\label{b1_rel}
\end{eqnarray}
\end{widetext}

\noindent where $\left|\sigma\right \rangle$ is the last occupied state.

The $b_2$ coefficient must be changed also in accordance with (\ref{e_rel}). The final expression for $b_2$ reads:

\begin{equation}
b_2=\frac{1}{2 n_f^2} \left[1 + \frac{(Z \alpha)^2}{n_f^2}\left(\frac{n_f}{j_f+1/2}-\frac{3}{4}\right) \right] ,
\label{b2_rel}
\end{equation}

\noindent where $j_f$ is the total angular momentum quantum number of the last occupied state and $\alpha \approx 1/137$ is the fine structure constant. 

A summary of matrix elements $\left\langle k |V_{rel}|\lambda \right\rangle$ is given in the Appendix. Details on the derivation of $\left\langle \lambda|V_{rel}|\lambda \right\rangle$ can be found in Ref. [\onlinecite{Bransden}].

\section{Regularizing the perturbative series}
\label{regular}

Once the region near the anionic threshold and the large-$Z$ limit are described, one may try to find an interpolation between them. To this end, we use a regularization of the perturbative series, Eq. (\ref{Z_series}).\cite{renormS} The next two formal terms of the series

\begin{equation}
I_p=b_2 Z^2 + b_1 Z + b_0 + \frac{b_{-1}}{Z} ~,
\label{reg_series}
\end{equation}

\noindent coming, in principle, from higher order perturbative corrections, are instead used to force that, at $Z=N-1$, $I_p(N)=E_a(N-1)$ and $dI_p/dZ=s$. That is:

\begin{eqnarray}
E_a &=& b_2 (N-1)^2 + b_1 (N-1) + b_0 + \frac{b_{-1}}{(N-1)} ~, \nonumber \\ 
s &=& 2 b_2 (N-1) + b_1 - \frac{b_{-1}}{(N-1)^2} ~.
\label{requirements}
\end{eqnarray}

\noindent We get a linear system of two equations and two variables ($b_0$ and $b_{-1}$), yielding:

\begin{eqnarray}
b_0 &=& E_a - 3 b_2 (N-1)^2 - (2 b_1 - s) (N-1) ~, \nonumber \\ 
b_{-1} &=& 2 b_2 (N-1)^3 + (b_1-s) (N-1)^2 ~.
\label{sol_b}
\end{eqnarray}

\section{Detecting problematic points in the NIST database}

\begin{center}
\begin{figure}[htb]
\includegraphics[width=0.9\linewidth,angle=0]{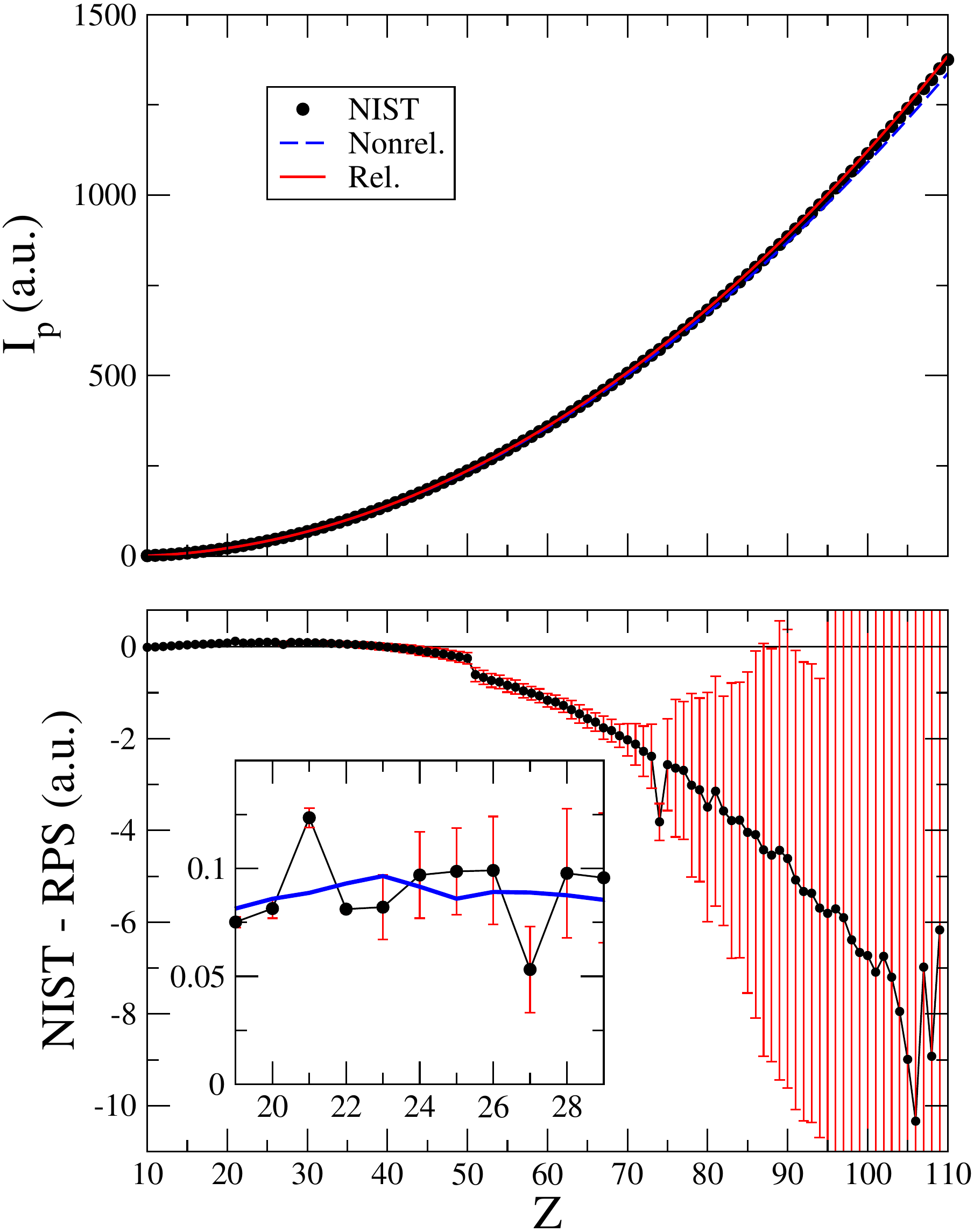}
\caption{\label{fig1} (Color online) The case of Ne-like ions ($N=10$). Upper panel: Ionization potential taken from the NIST compilation along with our nonrelativistic (discontinuous, blue) and relativistically corrected (continuous, red) RPS predictions versus atomic charge. Curves seem smooth at any scale. Lower panel: The difference between the NIST reported values and the RPS relativistic curve versus atomic charge. A 5-points running average curve (continuous, blue) is used as a reference for measuring deviations (see the inset). Inconsistencies are detected at $Z = 21$, 22, 27, 74 and $Z=50-51$. We stress also that there is a great dispersion of the data for $Z>100$, which is however within the hughe error bars reported for these points.}
\end{figure}
\end{center}

We would like to show how Eq. (\ref{reg_series}), with the coefficients $b_1$ and $b_2$ given in (\ref{b1_rel}) and (\ref{b2_rel}), respectively, and $b_0$ and $b_1$ coming from (\ref{sol_b}),  can be used to detect inconsistencies in the NIST data for the ionization potential of atomic ions. We study four isoelectronic systems (Fig. \ref{fig1}-\ref{fig3}) in quality of examples. An exhaustive revision will be published elsewhere.\cite{NISTrev1,NISTrev2}

In Figs. \ref{fig1}-\ref{fig3}, we plot the non-relativistic and the relativistic RPS, along with the NIST data, for these sequences. The lower panels show the difference NIST - RPS(relativ). The maximum relative errors are near 1\% for $N=10$, 2\% for $N=28,29$, and rises to around 10\% for $N=60$. In the first case, $N=10$, our approximate treatment of relativity does not reproduce the correct asymptotics at large $Z$. In the rest of the systems, both asymptotics ($Z=N-1$ and large $Z$) are correct, and the maximum errors are reached at intermediate $Z$, as it is common with interpolants \cite{renormS}. 

\begin{widetext}
\begin{center}
\begin{figure}[htb]
\includegraphics[width=0.9\linewidth,angle=0]{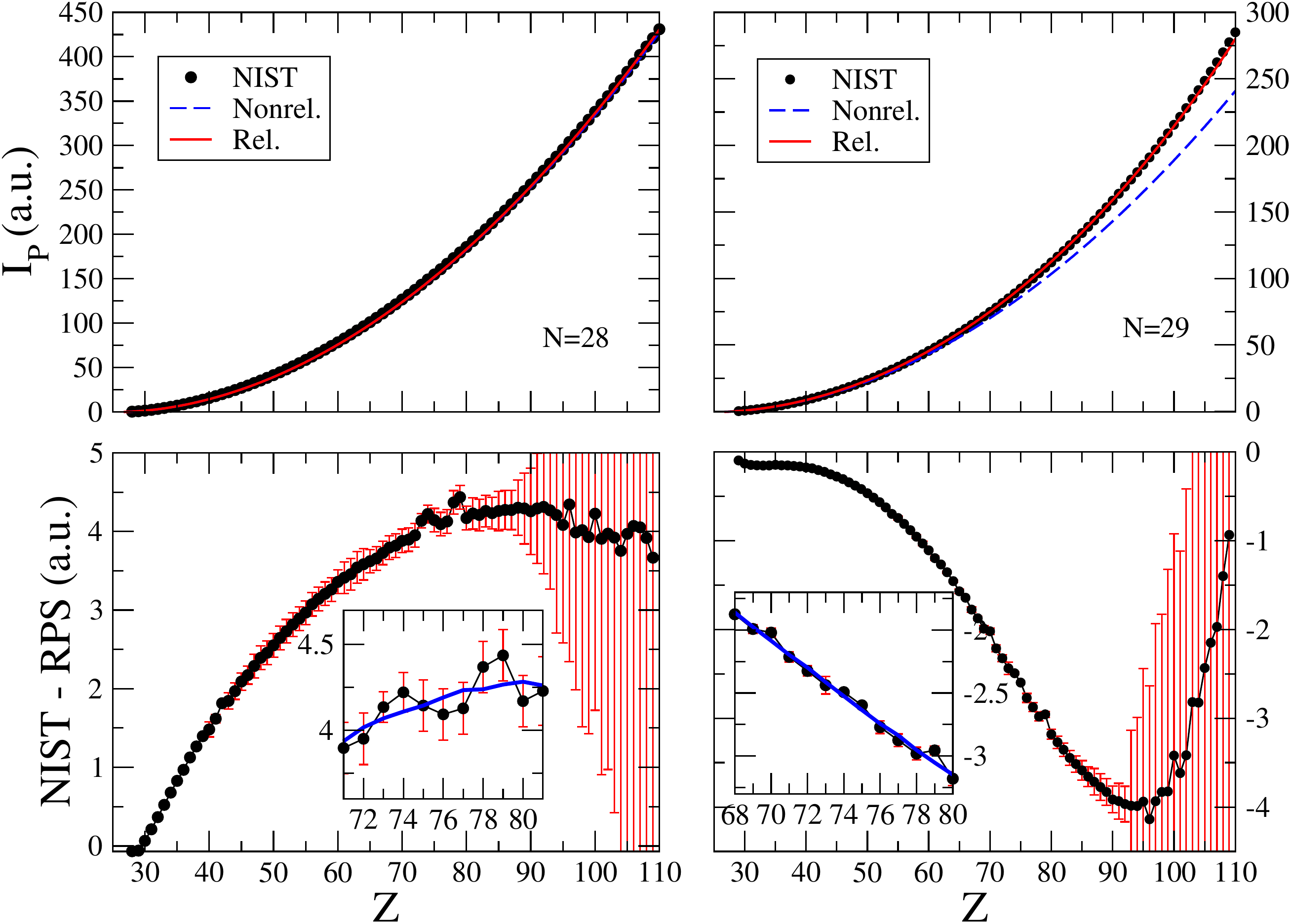}
\caption{\label{fig2} (Color online) The same as Fig. \ref{fig1} for Ni- ($N=28$, left panels) and Cu-like ($N=29$, right panels) ions. In the $N = 28$ system, we find inconsistencies at $Z=42$, 74, and 79, whereas for $N = 29$ small deviations at $Z=70$, and 79 are noticed. Relativistic effects are  significant for Cu-like ions at large $Z$ values, in which the last electron occupies a 4s orbital.}
\end{figure}
\end{center}
\end{widetext} 

We claim that, in spite of the fact that our relativistic RPS does not have spectroscopic precision, abrupt changes in the difference NIST - RPS may be a sign of inconsistency. Indeed, abrupt changes in $I_p$ are related to rearrangements of the electronic spectrum. In the interval between rearrangements or for large enough $Z$, the occupancy of orbitals is fixed, and $I_p$ should be smooth. The difference with our smooth RPS interpolant should also be a smooth function of $Z$.

Ne-like systems are closed shell, and do not exhibit rearrangements at any $Z$. In order to make evident inconsistent points in the NIST data, we construct an average NIST-RPS curve by means of a 5-points running average. In Fig. \ref{fig1}, the $Z=21$ point is so far from the average curve, for example, that it should be corrected. We can even give an estimate of the needed correction by measuring the distance to the average curve, which in this case is -0.034 a.u.

In Ni-like ions, the 3d$^{10}$ electronic configuration is reached already for $Z\ge 29$. Thus, we expect a smooth dependence from this point on. The $Z=74$ point, for example, is deviated from the average curve in 0.141 a.u., and its error bar is only 0.115 a.u. wide. Cu-like ions, on the other hand, show a 3d$^{10}$4s configuration at any $Z$. Nd-like ions experience rearrangements at various 
$Z$ values, but in the neighbourhood of $Z=74$, a problematic point, the difference should be smooth. Thus, we can undoubtedly distinguish this point.

\begin{center}
\begin{figure}[htb]
\includegraphics[width=0.9\linewidth,angle=0]{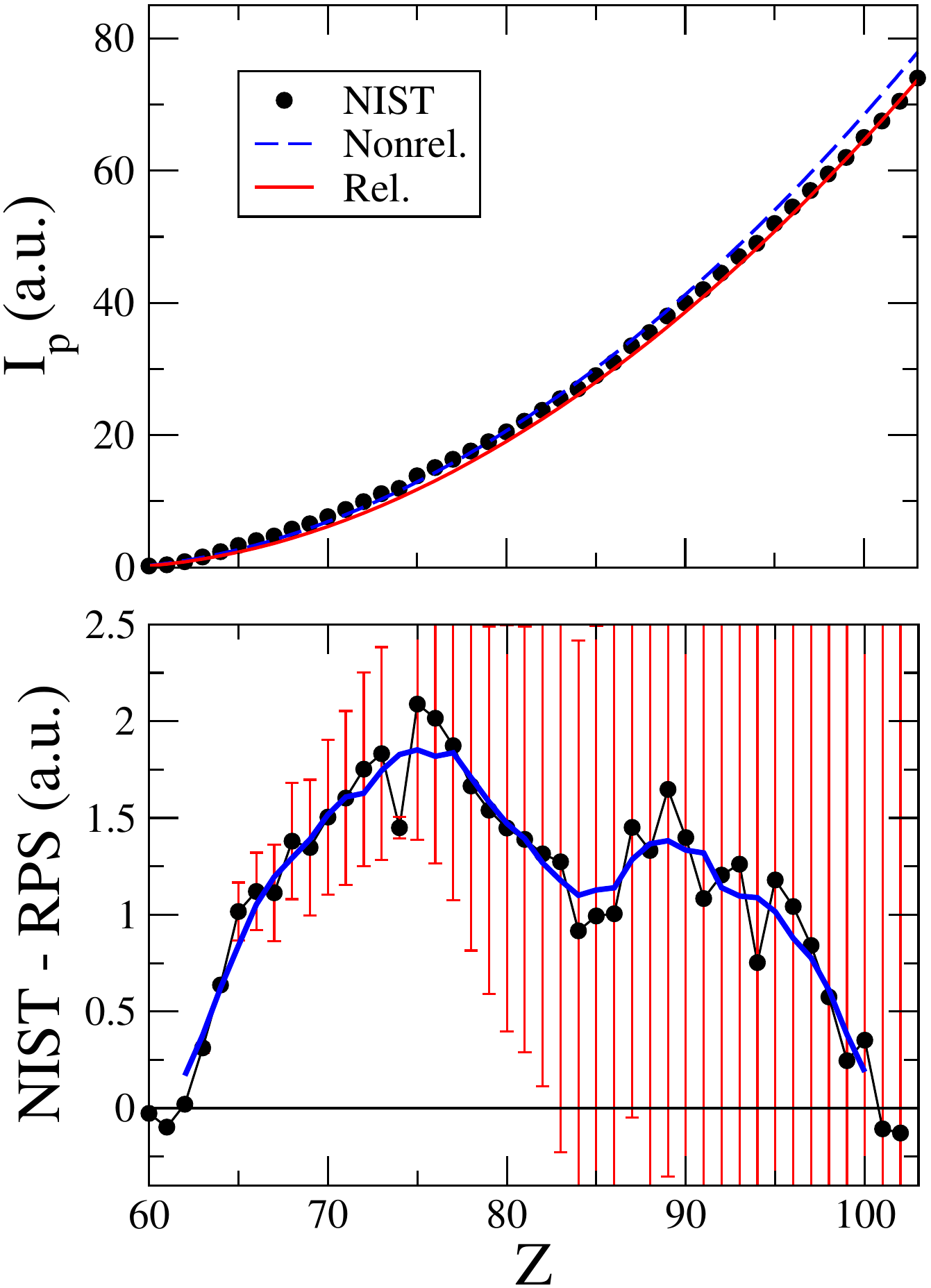}
\caption{\label{fig3} (Color online) The same as Fig. \ref{fig1} but for Nd-like ions ($N=60$). Only the $Z=74$ point is clearly inconsistent.}
\end{figure}
\end{center} 

\section{Conclusions}

The main result of the present paper is an analytical expression for the ionization energies of $N$-electron ions ($Z \geq N-1$) based on first-principles. This formula is exact in the $Z=N-1$ and large $Z$ regions. In the transition region, the error is only a few percents of the total ionization potential.

We show that our RPS expression may help identifying problems in a large database, such as the NIST compilation. A detailed analysis of the data is presented in Refs \onlinecite{NISTrev1} and \onlinecite{NISTrev2}.

\begin{acknowledgments}
The authors are grateful to the Caribbean Network for Quantum Mechanics, Particles and Fields (ICTP) for support. G.G. also acknowledge financial support from the European Community's FP7 through the Marie Curie ITN-INDEX.
\end{acknowledgments}

\appendix
\section{Explicit matrix elements for relativistic corrections}
\label{matrix_elements}

We can write the relativistically corrected Hamiltonian as:

\begin{equation}
H = H_0 + V_{rel} ~,
\label{htotal}
\end{equation}

\noindent where the non-perturbed Hamiltonian is given by:

\begin{equation}
H_0=-\frac{1}{2}\nabla^2-\frac{Z}{r}=T+V ~,
\label{h0}
\end{equation}

\noindent and the perturbation is expressed as a sum of terms:

\begin{eqnarray}
V_{rel} &=& H_1 + H_2 + H_3 ~,\nonumber\\
H_1 &=& -\frac{\alpha^2}{2} T^2 =  -\frac{\alpha^2}{2} \left(H_0+\frac{Z}{r}\right)^2 ,\nonumber\\
H_2 &=& \frac{Z \alpha^2}{2}\frac{1}{r^3} \vec{L} \cdot \vec{S} = \frac{Z \alpha^2}{4}\frac{1}{r^3} (J^2 - L^2 - S^2) (1-\delta_{l 0}) ~,\nonumber\\
H_3 &=& \pi \frac{Z \alpha^2}{2} \delta(\vec{r}) \delta_{l 0} ~.
\label{vrel}
\end{eqnarray}

\noindent where $\alpha$ is the fine structure constant, and $l$ is the orbital angular momentum quantum number.

We consider first-order perturbative corrections due to $V_{rel}$. The relevant matrix element is $\left \langle i \right | V_{rel} \left | \lambda \right \rangle$, where $\left | i \right \rangle$ is an eigenstate of $H_0$, $L^2$, $L_z$, $S^2$, $S_z$, and $\left | \lambda \right \rangle$ an eigenstate of  $H_0$, $L^2$, $S^2$, $J^2$, $J_z$. Only states such that $\left \langle i | \lambda \right \rangle = 0$ enter Eq. (\ref{wf_rel}), thus we restrict ourselves to this case. We have:

\begin{widetext}
\begin{eqnarray}
\left \langle i \left | V_{rel} \right | \lambda \right \rangle &=& \frac{Z^4\alpha^2}{2} \left[ -\left \langle i \left | 1/r \right | \lambda \right \rangle (\epsilon_i + \epsilon_{\lambda}) -\left \langle i \left | 1/r^2 \right | \lambda \right \rangle \right . \nonumber \\
&+& \left \langle i \left | 1/r^3 \right | \lambda \right \rangle \frac{1}{2} \left (j(j+1)-l(l+1)-\frac{3}{4} \right) + \pi \left \langle i \left | \delta(\vec{r}) \right | \lambda \right \rangle  \left. \right] ,
\label{vrel_elem}
\end{eqnarray}
\end{widetext}

\noindent where $j$ is the total angular momentum quantum number, and $\left | i \right \rangle$, $\left |\lambda \right \rangle$ are eigenstates of the Hydrogen ($Z=1$) Hamiltonian. $\epsilon_i$ and $\epsilon_\lambda$ are also scaled energies.  

Notice that $\left |\lambda \right \rangle$ states can be labelled by $n$ (principal quantum number), $l$, $j$ and $m_j$ (total angular momentum projection on $z$), whereas for $\left |i \right \rangle$ we need $n$, $l$, $m$ (orbital angular momentum projection on $z$), and $s_z$ (spin angular momentum projection on $z$). We can expand $\left | n l j m_j \right \rangle$ in terms of $\left | n l m s_z \right \rangle$ by means of the Clebsch-Gordan coefficients:

\begin{equation}
\left | n l s j m_j \right \rangle = \sum_{s_z, m} \left | n l s m s_z \right \rangle \left \langle  l s m s_z | j m_j \right \rangle ,
\label{cg_expansion}
\end{equation}

\noindent $\left \langle  l s m s_z | j m_j \right \rangle$ are non-vanishing only for $\left| l-s \right| \leq j \leq l+s$, and $m_j=m + s_z$.

The matrix elements entering Eq. (\ref{vrel_elem}) are explicitly written as:

\begin{widetext}
\begin{eqnarray}
\left \langle i \left | \delta(\vec{r}) \right | j \right \rangle &=& \delta_{l 0} \delta_{l l^{'}} \delta_{m m^{'}} \delta_{s_z s_z^{'}} \frac{1}{\pi} \frac{1}{\left(n n^{'}\right)^{3/2}} ~, \\
\left \langle i \left | 1/r^q \right | j \right \rangle &=& Z^q \delta_{l l^{'}} \delta_{m m^{'}} 
\delta_{s_z s_z^{'}} \theta(n^{'}-1-l)  2^{2l+2} \frac{\left(n n^{'}\right)^{l-q+1}}{\left(n+n^{'}\right)^{2l+2-q+1}} \sqrt{(n-l-1)! (n^{'}-l-1)! (n+l)! (n^{'}+l)!} \nonumber \\
&\times& \sum_{k=0}^{n-l-1} \sum_{k^{'}=0}^{n^{'}-l-1} \frac{ (-2)^{k+k^{'}} \frac{n^{k^{'}} {n^{'}}^{k}}{(n+n^{'})^{k+k^{'}}} \Gamma(k+k^{'}+2l+2-q+1) } { k! k^{'}! (n-l-k-1)! (n^{'}-l-k^{'}-1)! (2l+k+1)! (2l+k^{'}+1)!} ~.
\label{coul_elem}
\end{eqnarray}
\end{widetext}

\end{document}